\documentclass[%
 reprint,
showpacs,preprintnumbers,
 amsmath,amssymb,
 aps,
prb,
]{revtex4-1}

\usepackage{graphicx}
\usepackage{dcolumn}
\usepackage{bm}

\begin{document}

\title{Electronic structure of charged bilayer and trilayer phosphorene}

\author{Bukyoung Jhun}
\author{Cheol-Hwan Park}
\email{cheolhwan@snu.ac.kr}
\affiliation{Department of Physics, Seoul National University, Seoul 08826, Korea}

\date{\today}

\begin{abstract}
We have investigated the electronic structure of charged bilayer and trilayer
phoshporene using first-principles, density-functional-theory calculations.
We find that the effective dielectric constant for an external
electric field applied perpendicular to phosphorene layers
increases with the charge density and is twice as large as
in an undoped system if the electron density
is around $5\times10^{13}$\,cm$^{-2}$.
It is known that if few-layer phosphorene is placed under such an electric field,
the electron band gap decreases and if the
strength of the electric field is further increased, the band gap closes.
We show that the electronic screening due to added charge carriers reduces the amount of this
reduction in the band gap and increases the critical strength of the electric field for gap closure. If the electron density is around
$4\times10^{13}$\,cm$^{-2}$, for example,
this critical field for trilayer phosphorene is 40\,\% higher than that for a
charge-neutral system. The results are directly relevant to experiments on
few-layer phosphorene with top and bottom electrical gates and / or with chemical dopants.
\end{abstract}
\maketitle


\begin{figure}
\includegraphics[width=1\columnwidth]{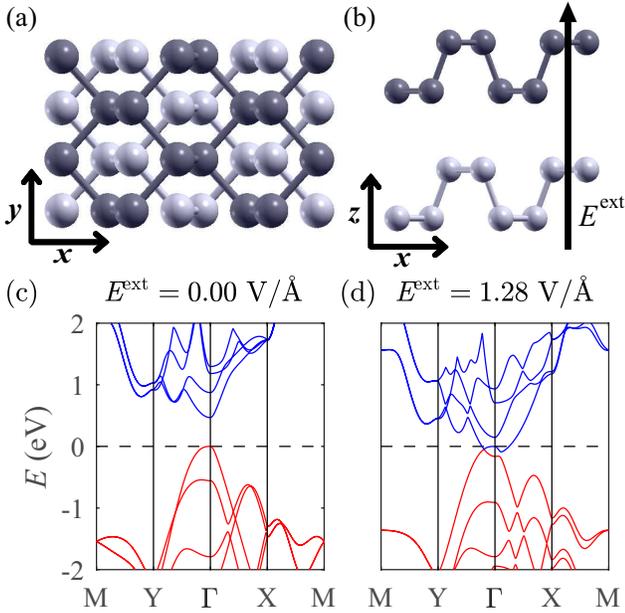}
\caption{(a) and (b) Atomic structures of bilayer phosphorene.
(c) and (d) Electronic band structures of bilayer phosphorene
without [(c)] and with [(d)] an external electric field.}
\label{fig_structure}
\end{figure}

Phosphorene, a single-layer of black phosphorus, has recently been
isolated by mechanical exfoliation~\cite{Li:2014, 2014ApPhL.104j3106K, Liu:2014}.
Among many exfoliated two-dimensional materials, phosphorene is unique in that due to its
structural anisotropy [Fig.~\ref{fig_structure}(a)] the thermal and electrical
transport~\cite{Liu:2014,Fei:NanoLett}, critical strains~\cite{Wei:2014}
and Young's moduli~\cite{Wang:2015, Jiang:2014}
along the two in-plane directions [$x$ and $y$ in Fig.~\ref{fig_structure}(a)] are different.

The electronic structure of few-layer phosphorene has been tuned
by chemical doping that effectively acts as an electric field
along the surface-normal direction and at the same time
as additional charge carriers~\cite{Kim:2015} [Fig.~\ref{fig_structure}(b)].
Also, a recent experimental study shows that top and bottom electrical gating
can tune the band gap of few-layer phosphorene efficiently~\cite{Deng:NatComm}.

There have been several density-functional-theory studies on few-layer phosphorene
since the first such studies~\cite{Li:2014,2014ApPhL.104j3106K,Liu:2014}.
Two-dimensional massless Dirac fermions are generated
if few-layer phosphorene is placed under a
strong enough external electric field
irrespective of whether the spin-orbit coupling
is taken into account or not~\cite{Baik:2015}.
How the electronic structure changes with the number of layers~\cite{Kumar:2016,Dolui:2015},
stacking order~\cite{Dai:2014, Cakir:2015}, strength of an applied electric
field~\cite{Baik:2015,Dai:2014,Kumar:2016, Liu:2015,PhysRevB.94.205426,Dolui:2015}
and strain~\cite{Peng:2014} has been extensively studied.
The effects of various transition metal dopants have also been
investigated~\cite{Zhang:2015a, Yu:2015, Sui:2015, Jing:2015, Hu:2015}.
The possibility of electron-doped phosphorene being a superconductor
above liquid-helium temperature was considered~\cite{Shao:2014}.

Although there have been many such first-principles-based
density-functional-theory studies on few-layer phosphorene,
the joint effects on the electronic structure of an external electric field
and added charge carriers while controlling the two
factors independently have not
been investigated so far. However, such a computational study
is directly relevant to the dual-electrical-gate (top and bottom gates)
experiments on few-layer phosphorene either with or without molecular dopants
and to device applications based on few-layer phosphorene.
(A study on bilayer graphene in this spirit was
performed~\cite{PhysRevB.79.165431}.)

In this paper, the effects of doping with charge carriers
on the effective dielectric constant
[for a field along $z$ as shown in Fig.~\ref{fig_structure}(b)]
and on the field-induced band gap modulation and Lifshitz phase transitions
of bilayer and trilayer phosphorene are studied from
first-principles-based density-functional-theory calculations.
The screening from added charge carriers
reduces the net electric field at
the interlayer region. This reduction in the electric field results in an increase
in the
effective dielectric constant. For example, if the electron density is
$5\times10^{13}$\,cm$^{-2}$,
the effective dielectric constant is roughly twice as large as its value
in undoped systems.
Moreover, the band gap reduction due to an external electric field is decreased due to
enhanced screening from the doped charge carriers.
Hence, the strength of the critical external electric field for gap closure
[Figs.~\ref{fig_structure}(c) and~\ref{fig_structure}(d)] is increased upon doping.
For example, the critical field for trilayer phosphorene
if the electron density is $4\times10^{13}$\,cm$^{-2}$ is 40\,\% higher than that for the
charge-neutral system.


Our first-principles calculations were performed with Quantum Espresso
package~\cite{Giannozzi:2009}. We used a plane-wave basis set with
a kinetic-energy cutoff
of 50~Ry and adopted the generalized gradient approximation of Perdew, Burke, and
Ernzerhof~\cite{PhysRevLett.77.3865}
for exchange correlation energy. Core-valence interactions were modeled
by Troullier-Martins norm-conserving pseudopotentials~\cite{Troullier:1993}.
We have not considered spin-orbit interactions since the
screening properties are not affected much.
The Brillouin zone is sampled by a $100\times100\times1$ Monkhorst-Pack
grid~\citep{PhysRevB.13.5188}. Note that such a dense sampling
is necessary for convergence in the case of doped, metallic systems
contrary to the case of undoped, semiconducting system.

The stable structure is obtained by fully relaxing the atomic positions until
all the components of the Hellmann-Feynman force on each atom are
weaker than 0.01~eV/{\AA}. Van der Waals interactions
are accounted for by using an empirical correction method~\cite{Grimme:2006}.
This method has been proven to be effective in describing layered
materials~\cite{Grimme:2007, Antony:2008, Hu:2014}.
We have checked that the atomic displacements due to extra doped charges
or vertical external electric fields considered in our work are less than 0.5\,\%
of the bond lengths (since the material is not polar).
We have checked that in agreement with previous
studies~\cite{Zhang:2015b, Cakir:2015, Dai:2014}
the stable stacking order of bilayer
phosphorene is of AB type such that even layers are shifted half the lattice period
along $y$ with respect to the odd layers [Fig.~\ref{fig_structure}(a)].
Bulk black phosphorus is also of AB type~\cite{Asahina:1984}.
As in previous first-principles
studies~\cite{Baik:2015,Dai:2014, Liu:2015,PhysRevB.94.205426,Dolui:2015},
the band gap closes if the external electric field [Fig.~\ref{fig_structure}(b)]
is strong enough [Figs.~\ref{fig_structure}(c) and~\ref{fig_structure}(d)].


\begin{figure}
\centering
\includegraphics[width=1\columnwidth]{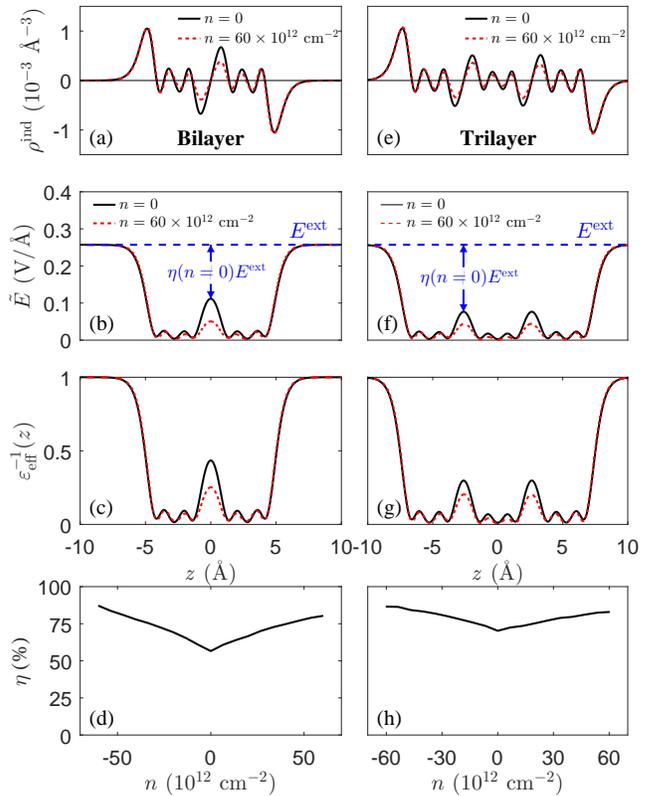}
\caption{(a) In-plane-averaged, induced electron density in bilayer phosphorene due to
an external electric field whose strength is $E^{\rm ext}=0.26$~V/\AA.
(b) The {\it external-field-induced} (in the sense that it vanishes if $E^{\rm ext}=0)$,
screened, and symmetrized electric field
$\tilde{E}(z)\equiv\frac{1}{(-e)}\cdot\frac{1}{2}\left[\left.-\frac{dV^{\rm ind}_{\rm tot}(z')}{dz'}\right|_{z'=z}-
\left.\frac{dV^{\rm ind}_{\rm tot}(z')}{dz'}\right|_{z'=-z}\right]$
in bilayer phosphorene. Here,
$V^{\rm ind}_{\rm tot}(z)$ is the
{\it external-electric-field-induced}, $xy$-plane-averaged total electrostatic
potential energy which is the sum of the external potential energy
$\left[=-(-e)\,z\,E^{\rm ext}\right]$ and {\it field-induced} Hartree potential energy.
(Note that the ionic potential energy does not change upon application of an external
electric field.)
The screening efficiency at the center of the interlayer region, $\eta$, is one and zero
if the screening is perfect and non-existent, respectively.
(c) The effective, position-dependent inverse dielectric function
$\varepsilon_{\rm eff}(z)^{-1}\equiv \tilde{E}(z)/E^{\rm ext}$ vs $z$.
In (a)-(c), solid black and dashed red curves are the results for undoped and
p-doped ($n=6\times10^{13}$~cm$^{-2}$) bilayer phosphorene, respectively.
(d) $\eta$ versus the charge density $n$. (We adopt the convention
that $n<0$ for electron doping and $n>0$ for hole doping.)
(e)-(h) Similar quantities as in (a)-(d) for trilayer phosphorene.}
\label{fig_induced_rho}
\end{figure}

An external electric field is modeled by
a saw-tooth type of potential with dipole-field
correction~\cite{PhysRevB.59.12301}
and a face-to-face distance between periodic images
is maintained to be 25--30~\AA.
The induced electron density due to an applied electric field
in both doped and undoped systems is
shown in Figs.~\ref{fig_induced_rho}(a) and~\ref{fig_induced_rho}(e).
Note that the additional charge carriers decrease the magnitude of the
field-induced charge density around
the interlayer regions. This decrease
naturally results in the reduction in the net, screened electric field at the interlayer
regions whereas in the other parts of the material the net electric field is not reduced
upon doping.

The calculated screening efficiency $\eta$ shown in Figs.~\ref{fig_induced_rho}(b), \ref{fig_induced_rho}(d),
\ref{fig_induced_rho}(f) and~\ref{fig_induced_rho}(h) (also, see caption) and effective, position-dependent
inverse dielectric function $\varepsilon_{\rm eff}^{-1}(z)$ shown in Figs.~\ref{fig_induced_rho}(c)
and~\ref{fig_induced_rho}(g) allow us to understand how the
effective dielectric constant increases with the amount of doped charge carriers.
In agreement with the induced charge density shown in Figs.~\ref{fig_induced_rho}(a) and~\ref{fig_induced_rho}(e),
the net effect of additional charge carriers is the reduction in the net, external-field-induced electric
field in the interlayer regions.

\begin{figure}
\includegraphics[width=1\columnwidth]{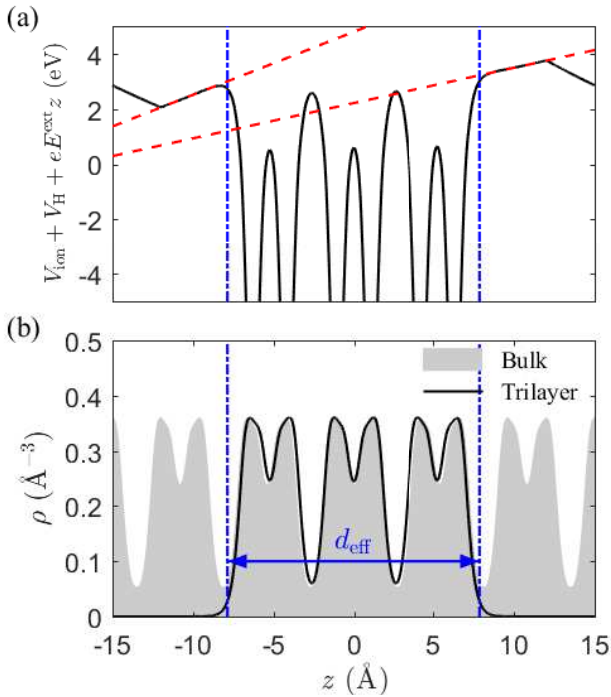}
\caption{(a) The sum of the ionic, Hartree and external potential
energies (or the total electrostatic potential energy)
averaged in the $xy$ plane versus $z$ for
n-doped trilayer phosphorene ($n=-4\times10^{13}$~cm$^{-2}$).
The slopes of the dashed red lines (divided by the charge $e$)
are the strengths of the electric fields in vacuum
and their average corresponds to $E^{\rm ext}=0.36$~V/\AA.
(b) The $xy$-plane-averaged electron densities of bulk (gray area) and
trilayer (solid black curve) phosphorene.
Vertical, dash-dotted blue lines indicate the positions where the electron density of
trilayer phosphorene is half the minimum electron density of the bulk crystal;
$d_{\rm eff}$ is the effective thickness of the slab.}
\label{fig_potential}
\end{figure}

The slopes of the electrostatic potential in the two vacuum regions
above and below the slab
differ from each other except for undoped case.
The average of the two slopes is equal to
the strength $E^{\rm ext}$ of the applied external electric field.
We determine the average of the {\it screened} electric field inside the
material $\left<E^{\rm scr}\right>$ 
by (i) obtaining the $xy$-plane-averaged
electrostatic potentials {\it linearly extrapolated} from the
two vacuum regions [as
illustrated by the two dashed red lines in Fig.~\ref{fig_potential}(a)]
at the (effective) top and bottom boundaries of the the slab
(we discuss how to choose these boundaries later),
(ii) taking their difference
and (iii) dividing this difference by the distance between those two effective
boundaries $(\equiv d_{\rm eff})$ [Fig.~\ref{fig_potential}(b)].

We need to define where the two {\it effective} boundaries of the material are.
The natural definition of the boundaries is the positions where the $xy$-plane-averaged
electron density is half the minimum value of the corresponding quantity
of the bulk crystal [Fig.~\ref{fig_potential}(b)].
The effective thickness $d_{\rm eff}$ defined in this way is almost the same as the bulk lattice
parameter along the $z$ direction times the number of layers.
The thicknesses of bilayer and trialyer phosphorene are 10.57~{\AA} and 15.49~{\AA},
respectively. The thicknesses per layer are 5.29~{\AA} and 5.16~{\AA}, respectively,
close to the corresponding bulk quantity, 5.3~{\AA}~\cite{Morita:1986}.

\begin{figure}
\includegraphics[width=1\columnwidth]{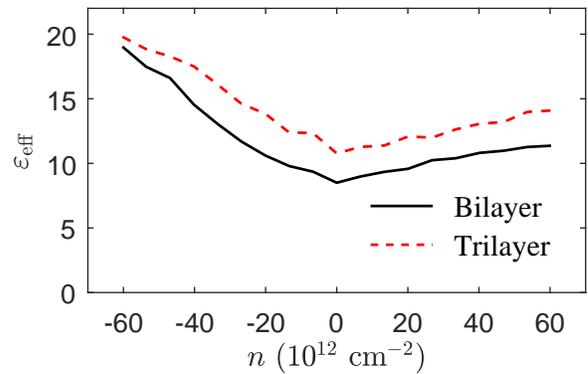}
\caption{Effective dielectric constant of bilayer (solid black curve) and
trilayer (dashed red curve) phosphorene versus the charge density.}
\label{fig_epsilon}
\end{figure}

Figure~\ref{fig_epsilon} shows the effective dielectric constant
$\varepsilon_{\rm eff}\equiv E^{\rm ext}/\left<E^{\rm scr}\right>$
of bilayer and trilayer phosphorene.
In our calculations, we used $E^{\rm ext}$ in the linear-response regime
for bilayer and trilayer phosphorene~\cite{Kumar:2016}
(0.15~eV/\AA$\le E^{\rm ext}\le0.36$~eV/\AA).
N-type doping is more efficient than p-type doping in screening the external electric field
and the effective dielectric constant is twice as large as
in undoped systems if the electron density is
around $5\times10^{13}$~cm$^{-2}$.

Calculated $\varepsilon_{\rm eff}$'s
of undoped ($n=0$ in Fig.~\ref{fig_epsilon})
bilayer and trilayer phosphorene are 8.5 and 10.8,
respectively.
Note that these values are much higher than those reported in
Ref.~\cite{Kumar:2016}: 2.9 for bilayer phosphorene and 3.5 for trilayer phosphorene.
This discrepancy arises from the difference in the definition of the effective
thickness $d_{\rm eff}$ [Fig.~\ref{fig_potential}(b)].
We found that $\varepsilon_{\rm eff}$ depends very sensitively
on $d_{\rm eff}$. In Ref.~\cite{Kumar:2016}, $d_{\rm eff}$ was defined as
the distance between the positions where the {\it induced} electron density
[see, e.\,g.\,, Figs.~\ref{fig_induced_rho}(a) and~\ref{fig_induced_rho}(e)] drops to $1\,\%$
of the nearest peak value. The values of $d_{\rm eff}$ of bilayer and trilayer phosphorene
defined in this way are 15.10~\AA\ and 20.06~\AA\ when $E_{\rm ext}=0.26$~eV/\AA,
respectively, which are longer than those
in this work by 43\% and 30\%, respectively. Our definition
of $d_{\rm eff}$ (i) does not require such an arbitrary criterion, (ii) can be defined
without considering an external electric field, and (iii) will be very relevant to device
applications since the effective thickness is defined based on the {\it total} electron density.

\begin{figure}
\includegraphics[width=1\columnwidth]{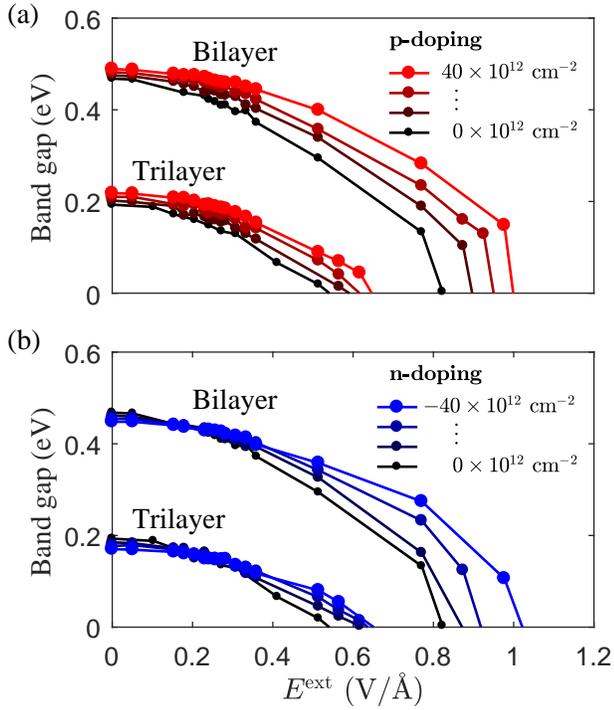}
\caption{Band gap of bilayer and trilayer phosphorene versus the
strength of the external electric field $E^{\rm ext}$. (a) and (b) show the results
for p-doped cases and n-doped cases, respectively.}
\label{fig_gap}
\end{figure}

\begin{figure}
\includegraphics[width=1\columnwidth]{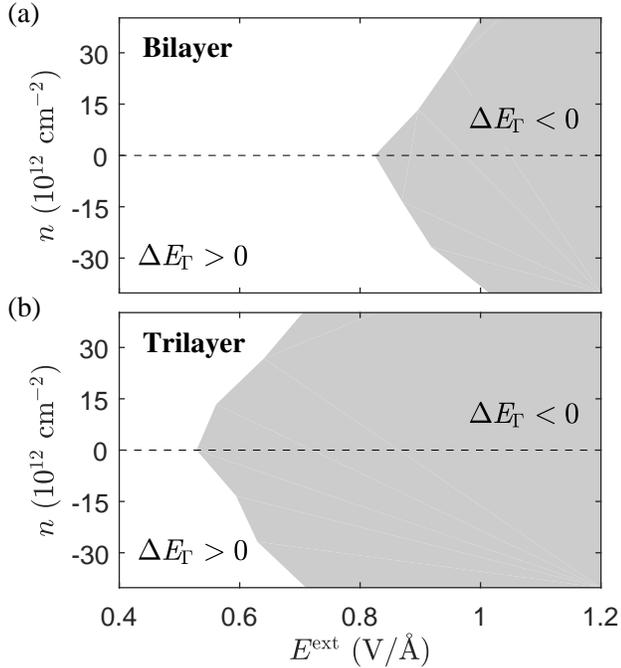}
\caption{Phase diagrams of bilayer [(a)] and trilayer [(b)] phosphorene.
The semiconducting phase with a positive energy gap at $\Gamma$ ($\Delta E_{\Gamma}>0$)
and the metallic phase ($\Delta E_{\Gamma}<0$) are shown in white and in gray, respectively.}
\label{fig_phase}
\end{figure}

Figure~\ref{fig_gap} shows how the band gap of bilayer and trilayer phosphorene varies upon doping and applying
an external electric field. If we neglect the small gap changes
at zero field due to added charge carriers,
the band gap decreases slower than in undoped system. Because of this behavior, the critical field strength
to close the band gap of few-layer phosphorene increases with the amount of
doped charge carriers (Fig.~\ref{fig_phase}).
For a doping with $n=4\times10^{13}$~cm$^{-2}$, the critical field
strength for bilayer and trilayer phosphorene
is increased by 20\,\% and 40\,\%, respectively.


In conclusion, we have shown that few-layer phosphorene doped with electrons or holes by
external electrical gates have different electronic and screening properties than an undoped
intrinsic counterpart. The effective dielectric constants can be doubled by
electrical gating and the strength of the critical electric field for gap closure
is significantly increased.
Our work also highlights the role of additional charge carriers
that can be easily introduced by electrical gates
in other few-layer two-dimensional materials.

\begin{acknowledgments}
This work was supported by the Creative-Pioneering Research Program through Seoul
National University.
\end{acknowledgments}

\bibliography{reference}

\end{document}